\documentclass[11pt,a4paper]{article}
\pdfoutput=1
\usepackage{jheppub}
\usepackage{amsmath}
\usepackage{bm}
\usepackage{amssymb}
\usepackage{color}

\def\slashchar#1{\setbox0=\hbox{$#1$}  
   \dimen0=\wd0     
   \setbox1=\hbox{/} \dimen1=\wd1  
   \ifdim\dimen0>\dimen1   
      \rlap{\hbox to \dimen0{\hfil/\hfil}} 
      #1     
   \else     
      \rlap{\hbox to \dimen1{\hfil$#1$\hfil}} 
      /      
   \fi}

\makeatletter
\def\overbracket#1{\mathop{\vbox{\ialign{##\crcr\noalign{\kern3\p@}
\downbracketfill\crcr\noalign{\kern3\p@\nointerlineskip}
$\hfil\displaystyle{#1}\hfil$\crcr}}}\limits}
\def\underbracket#1{\mathop{\vtop{\ialign{##\crcr
$\hfil\displaystyle{#1}\hfil$\crcr\noalign{\kern3\p@\nointerlineskip}
\upbracketfill\crcr\noalign{\kern3\p@}}}}\limits}
\def\upbracketfill{$\m@th\makesm@sh{\llap{\vrule\@height3\p@\@width.7\p@}}%
\leaders\vrule\@height.7\p@\hfill
\makesm@sh{\rlap{\vrule\@height3\p@\@width.7\p@}}$}
\def\downbracketfill{$\m@th
\makesm@sh{\llap{\vrule\@height.7\p@\@depth2.3\p@\@width.7\p@}}%
\leaders\vrule\@height.7\p@\hfill
\makesm@sh{\rlap{\vrule\@height.7\p@\@depth2.3\p@\@width.7\p@}}$}
\makeatother

\title{\boldmath Majorana dark matter in a classically scale invariant model}

\author[a]{Sanjin Beni\'c}
\author[a,b]{and Branimir Radov\v{c}i\'c}

\affiliation[a]{Department of Physics, University of Zagreb, Bijeni\v{c}ka c. 32, 10002 Zagreb, Croatia}
\affiliation[b]{Max-Planck-Institut f\"ur Kernphysik, Saupfercheckweg 1, 69117 Heidelberg, Germany}

\emailAdd{sanjinb@phy.hr}
\emailAdd{radovcic@mpi-hd.mpg.de}

\abstract{We analyze a classically scale invariant extension of the Standard Model with a dark gauge $U(1)_X$ broken by a doubly charge scalar $\Phi$ leaving a remnant $Z_2$ symmetry. Dark fermions are introduced as dark matter candidates and for anomaly reasons we introduce two chiral fermions. Due to classical scale invariance, bare mass term that would mix these two states is absent and they end up as stable Majorana fermions $N_1$ and $N_2$. We calculate cross sections for $N_aN_a \to \phi\phi$, $N_aN_a \to X^\mu \phi$ and $N_2N_2 \to N_1N_1$ annihilation channels. We put constraints to the model from the Higgs searches at the LHC, dark matter relic abundance and dark matter direct detection limits by LUX. The dark gauge boson plays a crucial role in the Coleman-Weinberg mechanism and has to be heavier than 680 GeV. The viable mass region for dark matter is from 470 GeV up to a few TeV. In the case when the two Majorana fermions have different masses, two dark matter signals at direct detection experiments could provide
a distinctive signature of this model.}

\begin{document}
\maketitle
\flushbottom

\section{Introduction}

With the recent discovery of the Higgs boson at the
LHC \cite{Aad:2012tfa,Chatrchyan:2012ufa} comes the hierarchy problem.
The null results in the first LHC run for supersymmetry, large extra dimensions and other
popular theoretical resolutions of the hierarchy problem invites us to think
about alternatives.
One of the possible alternatives is the classical scale invariance, first discussed by
Bardeen \cite{Bardeen:1995kv}. In this scenario, all masses are generated through the quantum breaking of scale invariance.
Such a mechanism is already present in Nature: in the
chiral limit the QCD part of the Standard Model (SM) Lagrangian
is scale invariant and the proton mass is determined by quantum effects.
Thus, the main idea is to promote classical
scale invariance to a general principle of Nature and apply it also to the electroweak (EW)
theory.

In perturbation theory, radiative mass generation can be
accomplished through the
Coleman-Weinberg (CW) mechanism \cite{Coleman:1973jx}
with the Higgs boson playing the role of the
scalon - the pseudo Goldstone
boson of the broken scale invariance \cite{Gildener:1976ih}.
Since the CW mechanism is not possible in the SM
due to the large top quark contribution,
a number of classical scale invariant
theories that go beyond the SM \cite{Meissner:2006zh,Foot:2007as,Espinosa:2007qk,AlexanderNunneley:2010nw},
in several directions \cite{Hempfling:1996ht,Chang:2007ki,Iso:2009ss,Englert:2013gz,Hambye:2013dgv,Carone:2013wla,Antipin:2013exa,Hashimoto:2013hta,Abel:2013mya,Hashimoto:2014ela,Kubo:2014ova,Kobakhidze:2014afa,Davoudiasl:2014pya,Kannike:2014mia,Lindner:2014oea,Hill:2014mqa,Allison:2014zya,Farzinnia:2014yqa,Khoze:2014xha,Allison:2014hna}
were presented.

Nowadays, there is an ample astronomical and cosmological evidence that around $27\%$ of the Universe is made of dark matter (DM). Its relic abundance points to the annihilation cross section of cold DM around the EW scale, making the Weakly Interacting Massive Particles (WIMPs) the leading DM candidates. This is another important hint for the physics beyond the SM which stimulates tremendous present and future
experimental efforts in the direct, indirect and accelerator searches for DM.

In a scale invariant setup any dimensionfull quantity arises from a common scale.
It is then tempting to establish a connection between the EW scale
and a priori unrelated DM scale by generating both through the CW mechanism.
This exciting possibility was investigated in a number of papers \cite{Hambye:2007vf,AlexanderNunneley:2010nw,Foot:2010av,Ishiwata:2011aa,Hur:2011sv,Heikinheimo:2013fta,Holthausen:2013ota,Farzinnia:2013pga,Gabrielli:2013hma,Radovcic:2014rea,Guo:2014bha,Farzinnia:2014xia,Kubo:2014ida,Altmannshofer:2014vra},
and is the main motivation for our work.
The model used here is described
by a dark $U(1)_X$ gauge group with a
doubly charged scalar $\Phi$ coupled to the SM via the Higgs portal.
In this construction the scale is generated through quantum corrections in the dark sector and
transmitted to the EW sector through the Higgs portal.
The dark sector contains two Majorana fermions $N_1$ and $N_2$.
Due to the remnant $Z_2$ symmetry both of these particles are
stable and are therefore DM candidates.

We extend the investigation of this model initiated in \cite{Radovcic:2014rea} in several directions.
We take into account that both Majorana fermions contribute to the DM relic abundance.
Besides the annihilation of $N_{1,2}$ into the $\phi\phi$ channel
we also include the $X^\mu\phi$ channel.
We use the recent LUX constraints for spin independent direct detection \cite{Akerib:2013tjd}.
Finally, we discuss two cases for the ratio of the Majorana fermions masses.
We find that the DM particle lies within the range of
470 GeV up to a few TeV.
The lower limit is set by the LHC constraints on the mixing
angle between the dark scalar and the SM scalar.
The upper limit comes from the estimated limitations
set by perturbation theory.
We find that the allowed parameter region of the model
is also constrained by the direct detection limits set by LUX.
The DM relic abundance is saturated by moderate
values of the dark Yukawas and dark gauge coupling.
The values for the spin independent cross section
that are presently allowed are in the reach of future experiments.
The sensitivity of the
forthcoming XENON1T \cite{Aprile:2012} and
LZ \cite{Mailing:2011} experiments
offers excellent perspective for testing this model.
In the case when the two Majorana fermions
have different masses, two signals at direct detection
experiments provide a distinctive signature for this model.

\section{The model}

The relevant part of the model Lagrangian is given as
\begin{equation}
\begin{split}
\mathcal{L} &=
i\bar{N}_L(\slashchar{\partial}-i g_X \slashchar{X}) N_L
- \frac{y_1}{2} \left(\bar{N}_L N_L^c  \Phi + \mathrm{h.~c.}\right)\\
&+i\bar{N}_R(\slashchar{\partial}-i g_X \slashchar{X}) N_R
- \frac{y_2}{2} \left(\bar{N}_R N_R^c  \Phi + \mathrm{h.~c.}\right)
\label{eq:lag}\\
&+|(\partial_\mu - 2ig_X X_\mu) \Phi|^2-\frac{1}{4}X_{\mu\nu}X^{\mu\nu}-
\frac{1}{2}\sin\epsilon B_{\mu \nu} X^{\mu\nu}- V(H,\Phi)~.
\end{split}
\end{equation}
where $N_{L,R}$ are left and right
chiral fields with the same charge
$g_X$ as required by anomaly cancellation, and
Yukawa couplings $y_{1,2}$ to the doubly $U(1)_X$ charged scalar $\Phi$.
The dark gauge $X$-boson couples to the $U(1)_Y$ SM
gauge field via the kinetic mixing parameter
$\sin\epsilon$ \cite{Holdom:1985ag}. This will lead to the coupling of the $X$-boson to a pair of SM particles
with the interactions suppressed by the mixing parameter $\sin\epsilon$ \cite{Chun:2010ve}.
After spontaneous symmetry breaking in the dark sector we end up with two Majorana fields
\begin{equation}
N_1 = N_L + N_L^c \, , \qquad N_2 = N_R + N_R^c~.
\end{equation}
The scalar potential is given as
\begin{equation}
V(H,\Phi) = \frac{\lambda_H}{2}(H^\dag H)^2+\frac{\lambda_\Phi}{2}(\Phi^\dag\Phi)^2+
\lambda_P(H^\dag H) (\Phi^\dag\Phi)~,
\label{eq:pot}
\end{equation}
where $H$ is the SM Higgs field, and $\Phi$ is a singlet under the SM group.

The Lagrangian in (\ref{eq:lag})-(\ref{eq:pot}) is classically
scale invariant.
Therefore all the masses in the model will be
generated dynamically.
This will be accomplished by a CW mechanism, which is a perturbative version of the QCD-like mass generation.
The breaking of scale symmetry is obtained by balancing the one-loop effective potential to its tree level counterpart.

We assume vacuum
expectation values (\emph{vevs}) for both scalars
\begin{equation}
H
=
\begin{pmatrix}
H^+\\
\frac{1}{\sqrt{2}}(v_H+h'+iG)\\
\end{pmatrix}
\, , \qquad \Phi = \frac{1}{\sqrt{2}}(v_\Phi+\phi'+iJ)~.
\end{equation}
In this work we use the GW framework \cite{Gildener:1976ih} which
imposes that the potential at the classical
level is flat along a particular direction in field space.
We first define a vector from scalar fields with non-vanishing \emph{vevs} given in the polar coordinates as
\begin{equation}
\begin{pmatrix}
v_H+h'\\
v_\Phi+\phi'\\
\end{pmatrix}
=r
\begin{pmatrix}
\sin\theta\\
\cos\theta\\
\end{pmatrix}
~.
\end{equation}
The rewritten tree-level potential reads
\begin{equation}
V(r) = r^4\left(\frac{\lambda_H}{8}\cos^4\theta + \frac{\lambda_\Phi}{8}\sin^4\theta+\frac{\lambda_P}{4}
\sin^2\theta\cos^2\theta\right)~.
\end{equation}
The vanishing bracket defines the
flat direction
\begin{equation}
\lambda_H(\Lambda)\lambda_\Phi(\Lambda)-\lambda_P^2(\Lambda)=0~,
\label{eq:flat}
\end{equation}
given at some common scale $\Lambda$ by the angle $\theta$
\begin{equation}
\sin^2\theta = -\frac{\lambda_P}{\lambda_H-\lambda_P}~.
\label{eq:cos}
\end{equation}
Minimizing the potential we can
also obtain the following relation between the \emph{vevs}
\begin{equation}
\frac{v_H^2}{v_\Phi^2} = - \frac{\lambda_P}{\lambda_H} ~.
\label{eq:vevs}
\end{equation}

Scalar mass eigenstates are given by
\begin{equation}
\begin{pmatrix}
h\\
\phi\\
\end{pmatrix}
=
\begin{pmatrix}
\cos\theta & -\sin\theta\\
\sin\theta & \cos\theta\\
\end{pmatrix}
\begin{pmatrix}
h'\\
\phi'\\
\end{pmatrix}~,
\end{equation}
where the mixing angle $\theta$ is given in (\ref{eq:cos}).
At the tree level the particle masses are
\begin{equation}
m_h^2 = (\lambda_H - \lambda_P) v_H^2~,
\end{equation}
\begin{equation}
m_X^2 = 4g_X^2 v_\Phi^2~,
\end{equation}
\begin{equation}
m_{N_a} = \frac{y_a}{\sqrt{2}}v_\Phi~,
\label{eq:dmmass}
\end{equation}
while $\phi$ remains massless at the tree level.

The common scale in the theory is generated by radiative corrections.
Then, the splitting in the spectra is solely due to the different
values of the dimensionless couplings in the model.
The quantum loop corrections are built on top
of the flat direction \cite{Gildener:1976ih}.
The one-loop scalar potential 
reads \cite{Gildener:1976ih,AlexanderNunneley:2010nw}
\begin{equation}
\delta V(r) = A r^4 + B r^4 \log\left(\frac{r^2}{\Lambda^2}\right)~,
\label{eq:one_loop}
\end{equation}
where \cite{AlexanderNunneley:2010nw,Radovcic:2014rea}
\begin{equation}
\begin{split}
A&=\frac{1}{64\pi^2 v_r^4}\Bigl\{m_h^4 \left(-\frac{3}{2}+\log\frac{m_h^2}{v_r^2}\right)+6m_W^4 \left(-\frac{5}{6}+\log\frac{m_W^2}{v_r^2}\right)\\
&+3m_Z^4 \left(-\frac{5}{6}+\log\frac{m_Z^2}{v_r^2}\right)+3m_X^4 \left(-\frac{5}{6}+\log\frac{m_X^2}{v_r^2}\right)\\
&-12m_t^4 \left(-1+\log\frac{m_t^2}{v_r^2}\right)-2m_{N_1}^4 \left(-1+\log\frac{m_{N_1}^2}{v_r^2}\right)\\
&-2m_{N_2}^4 \left(-1+\log\frac{m_{N_2}^2}{v_r^2}\right)\Bigr\}~,
\end{split}
\end{equation}
\begin{equation}
B = \frac{1}{64\pi^2 v_r^4}\left(m_H^4+6m_W^4+3m_Z^4+3m_X^4-12m_t^4-2m_{N_1}^4
-2m_{N_2}^4\right)~,
\end{equation}
and $v_r$ is the \emph{vev} of the field $r$.
The mass for the scalar $\phi$, the so-called scalon, is generated purely by radiative corrections.
In the GW framework \cite{Gildener:1976ih} this is
\begin{equation}
m_\phi^2 = \frac{1}{8\pi^2 v_H^2 \sin^2\theta}\left(m_H^4+6m_W^4+3m_Z^4+3m_X^4-12m_t^4-2m_{N_1}^4-2m_{N_2}^4\right)~.
\label{eq:scalmass}
\end{equation}

Due to the quantum numbers of $N_{L,R}$ a Dirac mass term
mixing the two components would be possible.
However, this would be in contradiction with the principle of
classical scale invariance.
Thus, both Majorana fermions are stable due to the respective remnant $Z_2$ symmetry making them the DM candidates in this model.
Related models of Majorana DM have been discussed recently.
In \cite{Garcia-Cely:2013nin} the heavier Majorana fermion is unstable, while in \cite{Cline:2014dwa} the mixing between $N_{L,R}$ was forbidden by a $Z_2$.
Due to the couplings to the SM particles via the kinetic mixing term in (\ref{eq:lag}) the dark X-boson is unstable.
For it to decay sufficiently quickly at the DM decoupling, the kinetic
mixing parameter takes on the value
larger than roughly $\sin\epsilon \gtrsim 10^{-8}$.
For possible $X$-boson LHC phenomenology, including a study of the
constraints on $\sin\epsilon$, see \cite{Chun:2010ve}.
LHC phenomenology of the scalar $\phi$ is similar to the scalar singlet extensions of the SM \cite{Farzinnia:2013pga,Farzinnia:2014xia,Altmannshofer:2014vra}.

\section{Annihilation cross sections}

\begin{figure}[h]
\centering
\centerline{\includegraphics[scale=0.5]{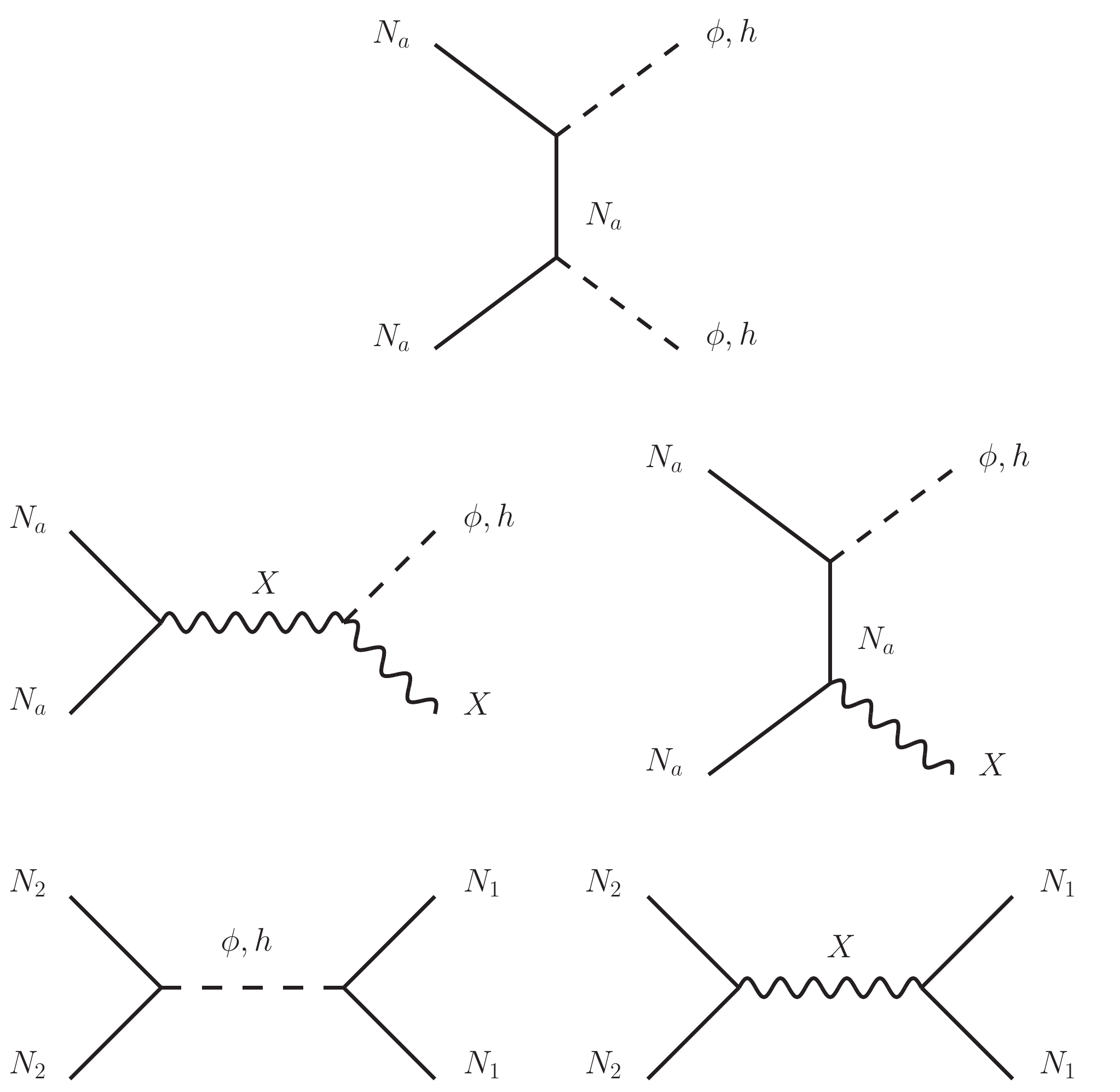}}
\caption{Annihilation processes of Majorana dark matter $N_1$ and
$N_2$ considered in this work.}
\label{fig:ann}
\end{figure}

In order to calculate the dark matter relic abundance we need to know the $N_{1,2}$ annihilation cross sections. In this work we will assume $m_{N_2}\geq m_{N_1}$ and consider the following channels
\begin{eqnarray*}
N_a N_a &\to& \phi\phi, \, h\phi, \, hh~,\\
N_a N_a &\to& X\phi , \, X h~,\\
N_2 N_2 &\to& N_1 N_1~,
\end{eqnarray*}
as shown on Fig.~\ref{fig:ann}.

The DM annihilation to SM particles is suppressed by a small portal coupling, the only communication between the dark and the SM sector.
We have checked that those channels don't play a numerically significant role.
We expand the cross sections in powers of the relative velocity $v$, and show only the dominant components, that is, the annihilation into the dark sector.

The cross section for the $N_a N_a \to \phi \phi$ channel is given as
\begin{equation}
\sigma (N_aN_a \to \phi\phi)v = \sigma^p_{aa\phi\phi}v^2 + \mathcal{O}(v^4)~,
\end{equation}
where
\begin{equation}
\sigma^p_{aa\phi\phi}=\frac{y_i^4 \cos^4\theta}{96\pi}
\frac{m_{N_a}^2(9m_{N_a}^4-8m_{N_a}^2 m_\phi^2 + 2m_\phi^4)}
{(m_\phi^2-2m_{N_a}^2)^4}\sqrt{1-\frac{m_\phi^2}{m_{N_a}^2}}~.
\end{equation}
The cross section for the $N_a N_a \to X \phi$ channel is given as
\begin{equation}
\sigma (N_aN_a \to X\phi)v = \sigma^s_{aa\phi X} + \sigma^p_{aa\phi X}v^2 + \mathcal{O}(v^4)~,
\end{equation}
where
\begin{equation}
\begin{split}
\sigma^s_{aa\phi X} &= \frac{g_X^6 \cos^2\theta}{16\pi}v_\Phi^2
\frac{m_X^4+(m_{\phi
}^2-4 m_{N_a}^2)^2-2 m_X^2 (m_{\phi }^2+4 m_{N_a}^2)}{m_{N_a}^2 m_X^6}\\
&\times
\sqrt{\frac{m_X^4+(m_{\phi
}^2-4 m_{N_a}^2)^2-2 m_X^2 (m_{\phi }^2+4 m_{N_a}^2)}{m_{N_a}^4}}
\end{split}
\end{equation}
and
\begin{equation}
\begin{split}
\sigma^p_{aa\phi X} &= \frac{g_X^2}{768\pi}\frac{\cos^2\theta}{m_{N_a}^2 m_X^6 (m_X^2-4m_{N_a}^2)^2(-4m_{N_a}^2+m_\phi^2+m_X^2)^4}\bigg\{
-8g_X^4 v_\Phi^2(-4m_{N_a}^2+m_\phi^2+m_X^2)^4\\
&\times\Big[384m_{N_a}^8-96 m_{N_a}^6(5m_\phi^2+7m_X^2)+8m_{N_a}^4(12m_\phi^2+6m_\phi^2m_X^2+43m_X^4)\\
&-2m_{N_a}^2(24m_\phi^4 m_X^2-37m_\phi^2 m_X^4+59m_X^6)+5m_X^4(m_\phi^2-m_X^2)^2\Big]\\
&-16\sqrt{2}y_a g_X^2v_\Phi m_{N_a} m_X^2(4m_{N_a}^2-m_X^2)(-4m_{N_a}^2+m_\phi^2+m_X^2)^2\\
&\times\Big[
128m_{N_a}^8-96m_{N_a}^6(m_\phi^2+m_X^2)+8m_{N_a}^4(3m_\phi^4+2m_\phi^2m_X^2+13m_X^4)\\
&-2m_N^2(m_\phi^2+m_X^2)(m_\phi^4-2m_\phi^2 m_X^2+9m_X^4)-m_X^4(m_\phi^2-m_X^2)^2\Big]\\
&+y_a^2 m_X^4(m_X^2-4m_{N_a}^2)^2\Big[m_X^{10}+4m_X^8(m_{N_a}^2-m_\phi^2)
+2m_X^6(16m_N^4+3m_\phi^4)\\
&-4m_X^4(m_\phi^2-4m_{N_a}^2)^2(10m_{N_a}^2+m_\phi^2)+m_{X}^2(m_\phi^2-4m_{N_a}^2)^2(80m_{N_a}^4+8m_{N_a}^2 m_\phi^2+m_\phi^4)\Big]\\
& +4m_{N_a}^2(m_\phi^2-4m_{N_a}^2)^4\bigg\}
\sqrt{\frac{m_X^4+(m_{\phi
}^2-4 m_{N_a}^2)^2-2 m_X^2 (m_{\phi }^2+4 m_{N_a}^2)}{m_{N_a}^4}}~.
\end{split}
\end{equation}
For simplicity and clarity we do not list the cross sections for annihilations to Higgs bosons, but we include them in our numerical calculations.
The cross section for the $N_2 N_2 \to N_1 N_1$ channel is given as
\begin{equation}
\sigma (N_2 N_2 \to N_1 N_1)v = \sigma^s_{2211} + \sigma^p_{2211} v^2 + \mathcal{O}(v^4)~,
\end{equation}
where
\begin{equation}
\sigma^s_{2211} = \frac{g_X^4}{4\pi}\frac{m_{N_1}^2}{m_X^4}
\sqrt{1-\frac{m_{N_1}^2}{m_{N_2}^2}}
\end{equation}
and
\begin{equation}
\begin{split}
\sigma^p_{2211}=&
\frac{1}{192\pi}\bigg\{
\frac{2 g_X^4}{(m_X^2-4m_{N_2}^2)^2(m_{N_2}^2-m_{N_1}^2)}
\Big[m_{N_1}^4(240m_{N_2}^4-120m_{N_2}^2 m_X^2+23m_X^4)\\
&-4m_{N_1}^2(48m_{N_2}^6-24m_{N_2}^4 m_X^2+7m_{N_2}^2m_X^4)+
8m_{N_2}^4 m_X^4\Big]\\
&+\frac{3y_1^2 y_2^2\cos^4\theta}{(m_\phi^2-4m_{N_2}^2)^2}(m_{N_2}^2-m_{N_1}^2)
\bigg\}\sqrt{1-\frac{m_{N_1}^2}{m_{N_2}^2}}~.
\end{split}
\end{equation}

\section{Dark matter constraints}

\subsection{Freeze-out and relic abundance}
\label{ssec:freeze}

In this model we have two DM candidates, $N_1$ and $N_2$.
If $N_1$ is much lighter than $N_2$, it could not annihilate to dark scalars and dark gauge bosons.
It could only annihilate directly to the SM particles, with too small cross sections to satisfy
the DM relic abundance constraint. Only possible solution would be for $N_1$ to decouple while
it is relativistic which would put strong bounds on its mass, around $m_{N_1} < 1$eV. That would
require a tiny Yukawa coupling $y_1$ and it would not be in line with the idea that all new particle masses are at the scale generated by quantum effects.
That is why we take the masses of $N_1$ and $N_2$ of the same order
of magnitude.

We will focus on two mass hierarchies, the degenerate mass case where $m_{N_2}=m_{N_1}$ and non-degenerate mass case with $m_{N_2} = 1.5 m_{N_1}$. In the degenerate case the annihilation of $N_1$ and $N_2$ into each other is not important for the DM relic abundance. In principle, with different masses, the channel $N_2 N_2 \to N_1 N_1$ mixes
the two coupled Boltzmann equations, but if the mass splitting is large enough, the DM freeze-out is sequential so that
the decoupling of $N_1$ happens after the decoupling of $N_2$.
Both particles leave the thermal bath at respective
temperatures $T_f^a$ given by $m_{N_a}/T_f^a=x_f^a \simeq 25$.
Lets set $\Delta x$ as the width of the freeze-out region.
Then the condition on $N_2-N_1$ mass ratio to have a sequential freeze-out is
\begin{equation}
\frac{m_{N_2}}{m_{N_1}} \gtrsim 1+\frac{\Delta x}{x_f^a}~.
\label{eq:dec}
\end{equation}
With typical values of $\Delta x\simeq 5$ and $x_f^a \simeq 25$
we have $m_{N_2}\gtrsim 1.2 m_{N_1}$.

The number density over entropy density today, $Y_\infty^a=n_{N_a}/s_0$ is
approximately given by the formula \cite{Griest:1988ma,Kolb:1990vq}
\begin{equation}
Y^a_\infty = \frac{3.79 x_f^a}{\sqrt{g_*} m_{Pl}m_{N_a} \left(\sigma^s_{aaX\phi}+\delta_{a2}\sigma^s_{2211}+\frac{3}{x_f^a}(\sigma^p_{aa\phi\phi}+\sigma^p_{aaX\phi}+\delta_{a2}\sigma^p_{2211})\right)}~,
\label{eq:freeze}
\end{equation}
where $g_* = 86.25$ is the number of relativistic degrees of freedom at $T_f$
and $m_{Pl}=1.22\times 10^{19}$ GeV is the Planck mass.
The observed dark matter relic abundance of the Universe is
\begin{equation}
\Omega_{DM} h^2 = \frac{\rho_{N_1}+\rho_{N_2}}{\rho_{cr}} = \frac{m_{N_1}n_{N_1}+m_{N_2}n_{N_2}}{\rho_{cr}}~.
\end{equation}

\subsection{Direct and indirect detection}
\label{ssec:dd}
\begin{figure}[h]
\centering
\centerline{\includegraphics[scale=0.5]{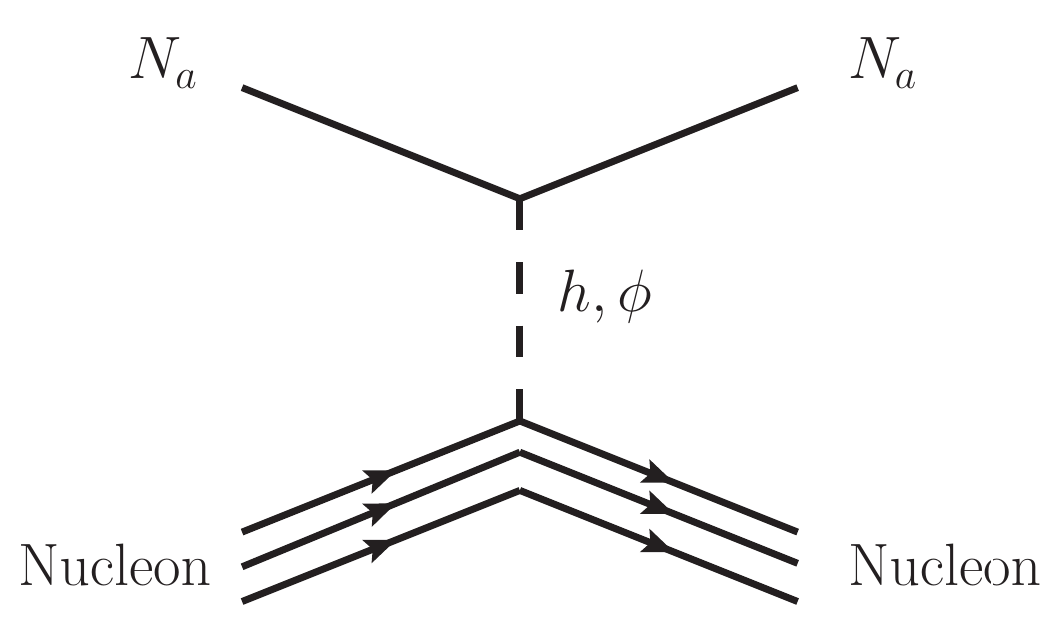}}
\caption{Scattering of Majorana dark matter $N_a$ of a nucleon.}
\label{fig:dir}
\end{figure}
It is possible that our DM candidates $N_1$ and $N_2$ are detected by elastic
scattering of a nucleus at the DM direct detection experiments.
In our model the interaction with the nucleon is governed
by the $t$-channel exchange of scalars $h$ and $\phi$ (see Fig.~\ref{fig:dir})
The spin independent cross section for direct detection is
given by \cite{Cerdeno:2010dd}
\begin{equation}
\sigma_{SI} = \frac{y_a^2}{2\pi}
\frac{m_p^4}{v_H^2} f^2 \sin^2\theta \cos^2\theta\left(\frac{1}{m_\phi^2}-\frac{1}{m_h^2}\right)^2~,
\label{eq:ddmaj}
\end{equation}
where $m_p$ is the proton mass and $f=0.35$ \cite{Cline:2013gha} parameterizes the nuclear matrix element.

We use the LUX published limits \cite{Akerib:2013tjd} on
the direct detection cross section.
We take into account the fact that we have two DM particles
contributing to the relic abundance with different
number densities and therefore different
event rates.
The modified LUX constraint scales as the ratio of the
number density of the particular DM component, $n_a$, and the number density for which
this component would saturate the DM relic abundance.

DM annihilations in the high density regions of the Universe could lead to indirect detection signals \cite{Cirelli:2010xx,Conrad:2014tla}.
In our case, $N_1$ and $N_2$ cannot annihilate directly into photons, so they cannot produce gamma lines.
If the positron excess measured by PAMELA and AMS-02 comes from DM annihilations, the cross section for DM annihilations
has to be larger than in the usual WIMP scenario \cite{Conrad:2014tla}.
Thus, our model cannot explain the positron excess measured by PAMELA or AMS-02.
In principle, DM annihilations will produce fluxes of SM particles, including the flux of continuum gamma rays
and antiprotons. However, searches for all such annihilations products are not
yet sensitive enough to reach the typical values of the WIMP cross section for DM masses above 1 TeV \cite{Conrad:2014tla} as found in our model (see the following Section).
Similar conclusions were reached in \cite{Hambye:2013dgv}.
We conclude that at the moment, our model is not constrained by present DM indirect searches.

\section{Results}

In the following we show the predictions of this model
for several quantities of interest.
There are six new parameters in the model: the gauge coupling $g_X$,
the two Yukawa couplings $y_1$ and $y_2$, and the three quartic
scalar couplings $\lambda_P$, $\lambda_H$ and $\lambda_\Phi$.
These are to be constrained by the Higgs boson mass
$m_h=125$ GeV and \emph{vev} $v_H =246$ GeV and the
DM relic abundance $\Omega_{DM} h^2 = 0.1187(17)$ \cite{Ade:2013zuv} leaving three undetermined parameters.

We will show our results by fixing the mass ratio of the Majorana
fermions (or, equivalently, the ratio of  $y_a$'s)
and varying the scalon $\phi$ and the dark $X$-boson mass in a
region set as follows.
A light scalon $\phi$ is
excluded in our model
by the LEP Higgs searches providing a lower bound
of $114$ GeV.
In addition to the large top quark contribution to the radiatively
generated scalon mass $m_\phi$ in (\ref{eq:scalmass}), there are also
contributions from DM Majorana fermions $N_1$ and $N_2$
that need to be overcome by the dark $X$-boson contributions
for the CW mechanism to work. This sets a lower limit on dark $X$-boson mass.
The parameter space we will cover starts from the initial values $m_\phi = 114$ GeV
and $m_X = 600$ GeV.

Our results also take into account the current LHC
data providing an upper bound on the scalar
mixing angle $\sin\theta<0.37$ \cite{Farzinnia:2013pga,Farzinnia:2014xia,Altmannshofer:2014vra}.
In our calculations we will also include the direct detection constraint
discussed in the previous Section.

\subsection{The case of degenerate masses}

\begin{figure}[h]
\centering
\centerline{\includegraphics[scale=0.54]{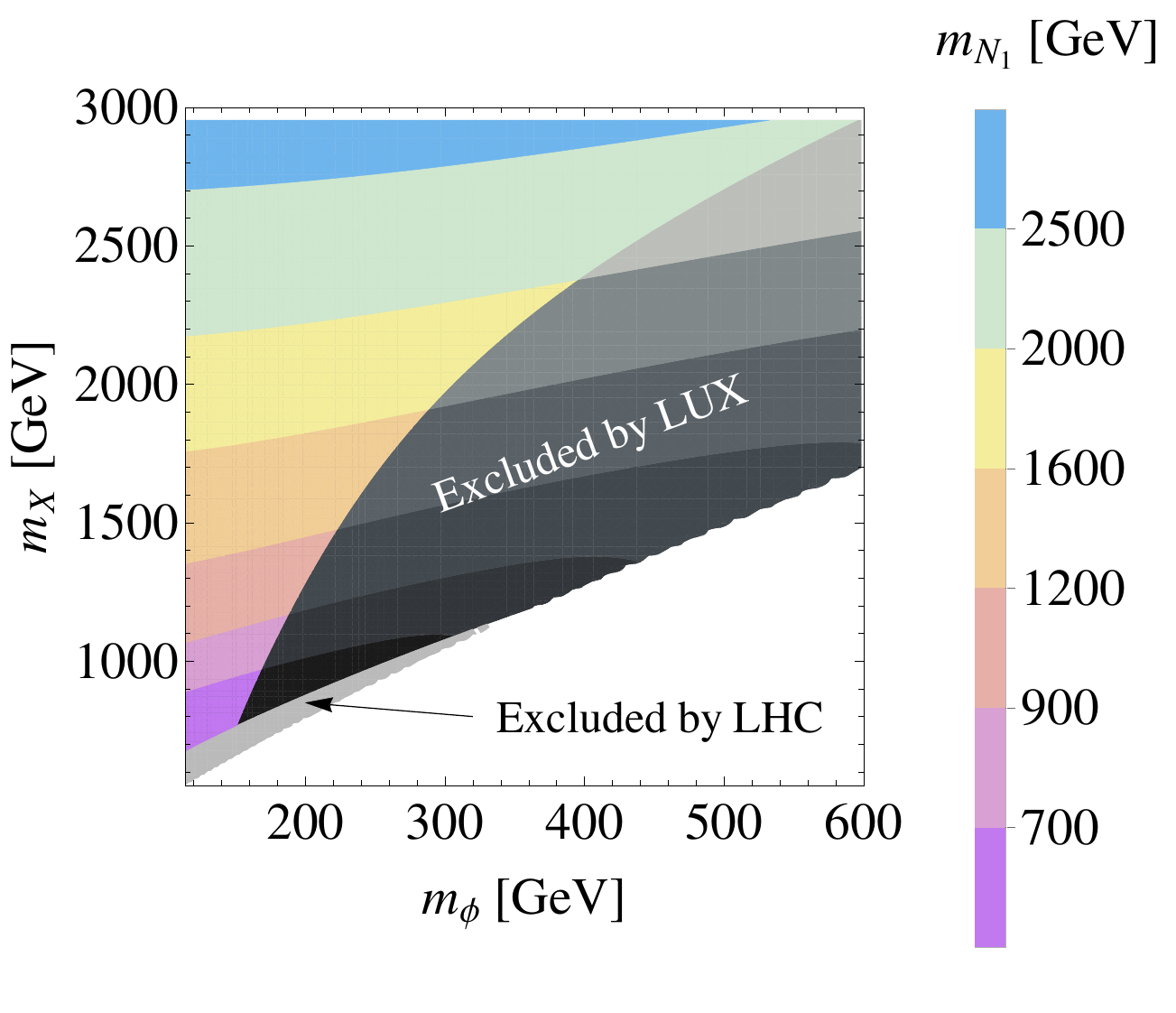}
\hspace{0.25cm}
\includegraphics[scale=0.53]{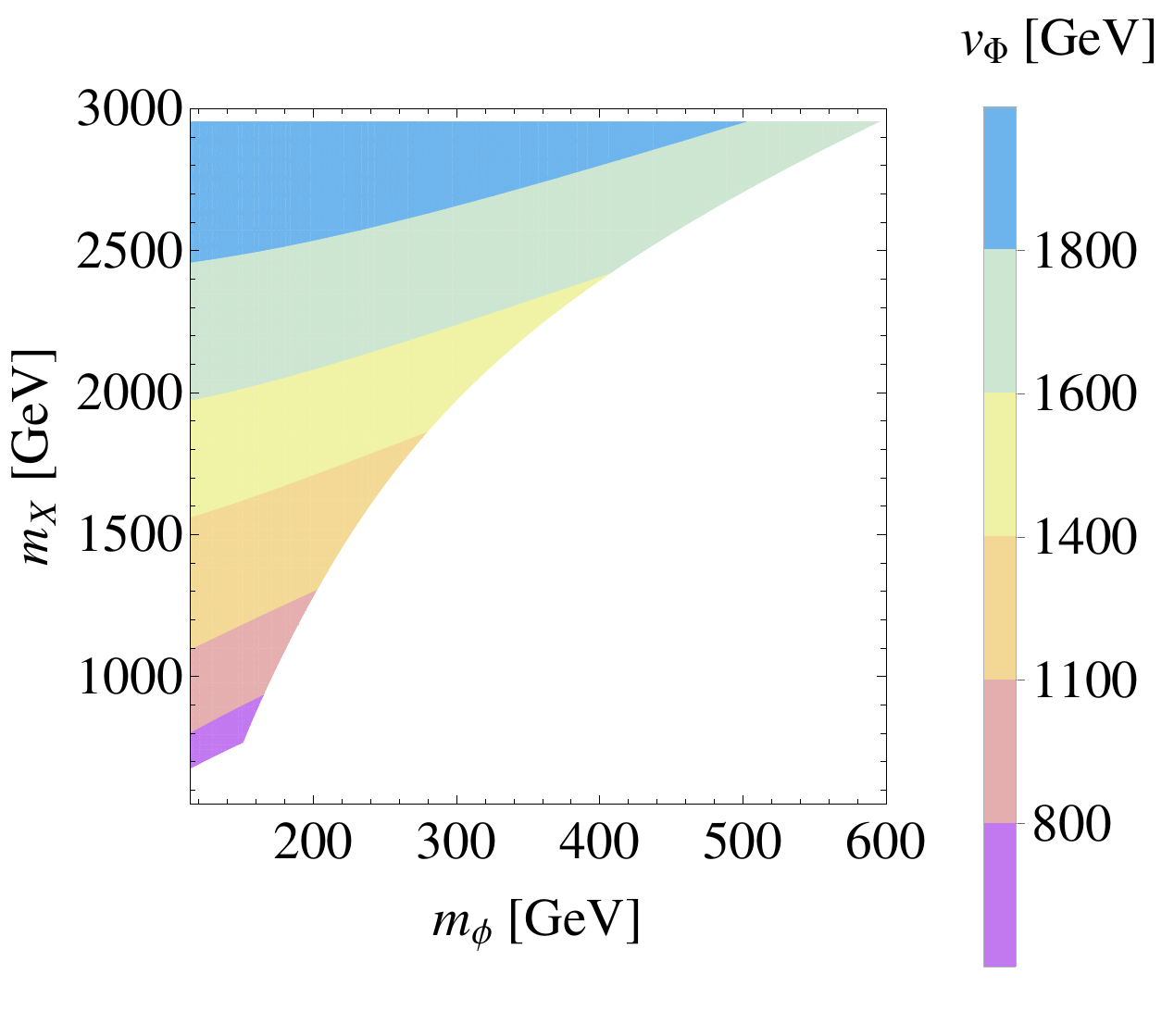}}
\caption{Left (right) panel: Mass of the DM Majorana
fermion $m_{N_1}$ (\emph{vev} of the $U(1)_X$ charged
scalar $v_\Phi$) in the $m_\phi-m_X$ plane for the degenerate mass case.
We impose the LHC bound on $\sin\theta$ and the LUX experiment constraints.}
\label{fig:mngx1}
\end{figure}

Here we focus on the scenario
$$m_{N_1} = m_{N_2}~,$$
where
both Majorana fermions contribute equally
to the observed DM relic abundance.
The main result is the Majorana
DM mass shown on the left panel
of Fig.~\ref{fig:mngx1} in the $m_\phi-m_X$ plane.
The region of low masses ($m_{N_1}\lesssim 470$ GeV)
is excluded by the current LHC bound on the scalar mixing
angle $\sin\theta <0.37$.
This also sets the lower bound on the dark $X$-boson to be $680$ GeV.
Due to its cardinal role in the CW mechanism, the dark gauge
boson is the heaviest particle in the dark sector.
A substantial portion of the parameter space has been excluded by
the LUX experiment \cite{Akerib:2013tjd} as depicted by
the gray scaled region
on the left panel of Fig.~\ref{fig:mngx1}.
The mass of the Majorana DM candidates
$N_1$ and $N_2$, is from $470$ GeV to a few TeV.

\begin{figure}[h]
\centering
\centerline{\includegraphics[scale=0.54]{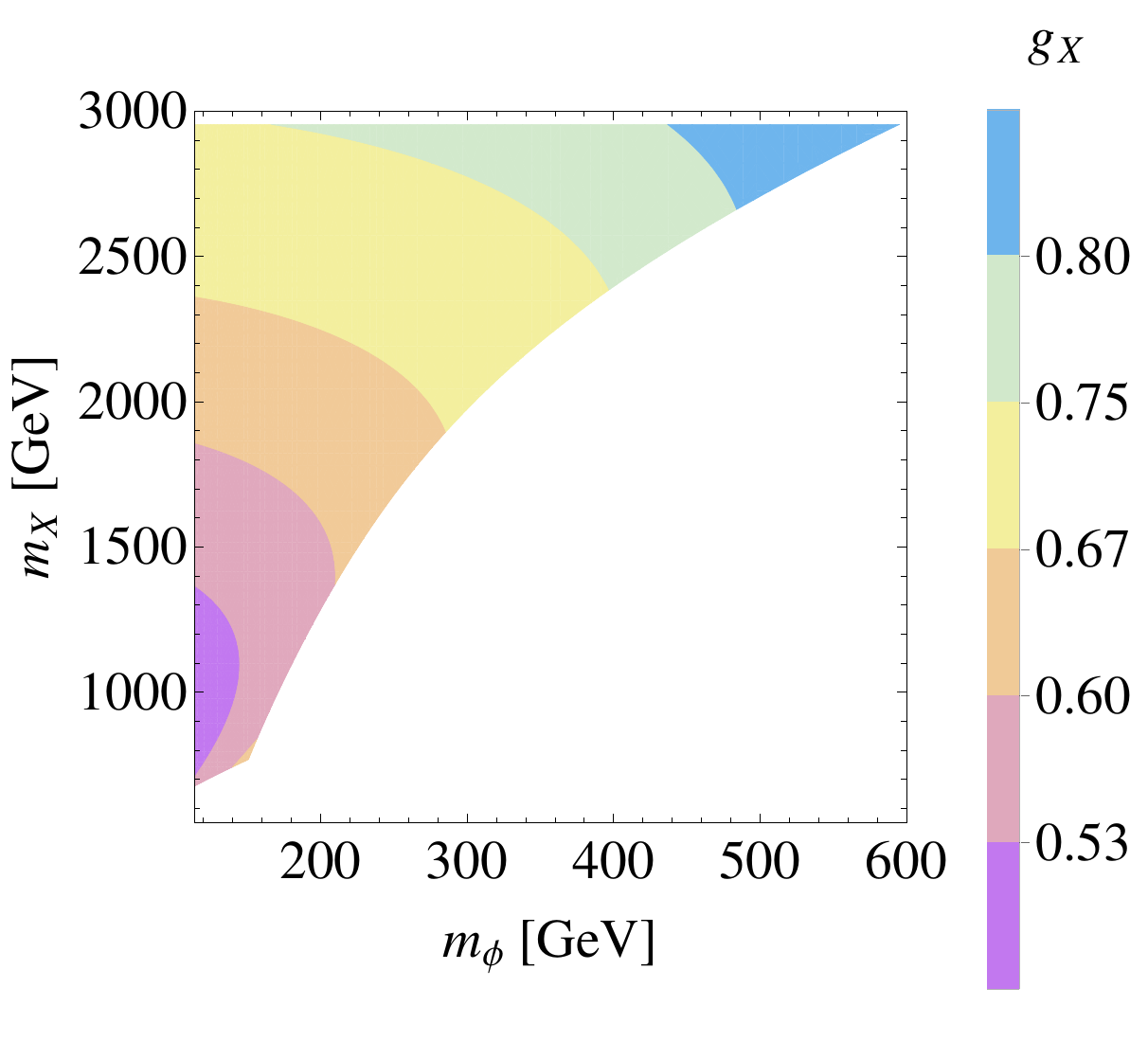}
\hspace{0.25cm}
\includegraphics[scale=0.53]{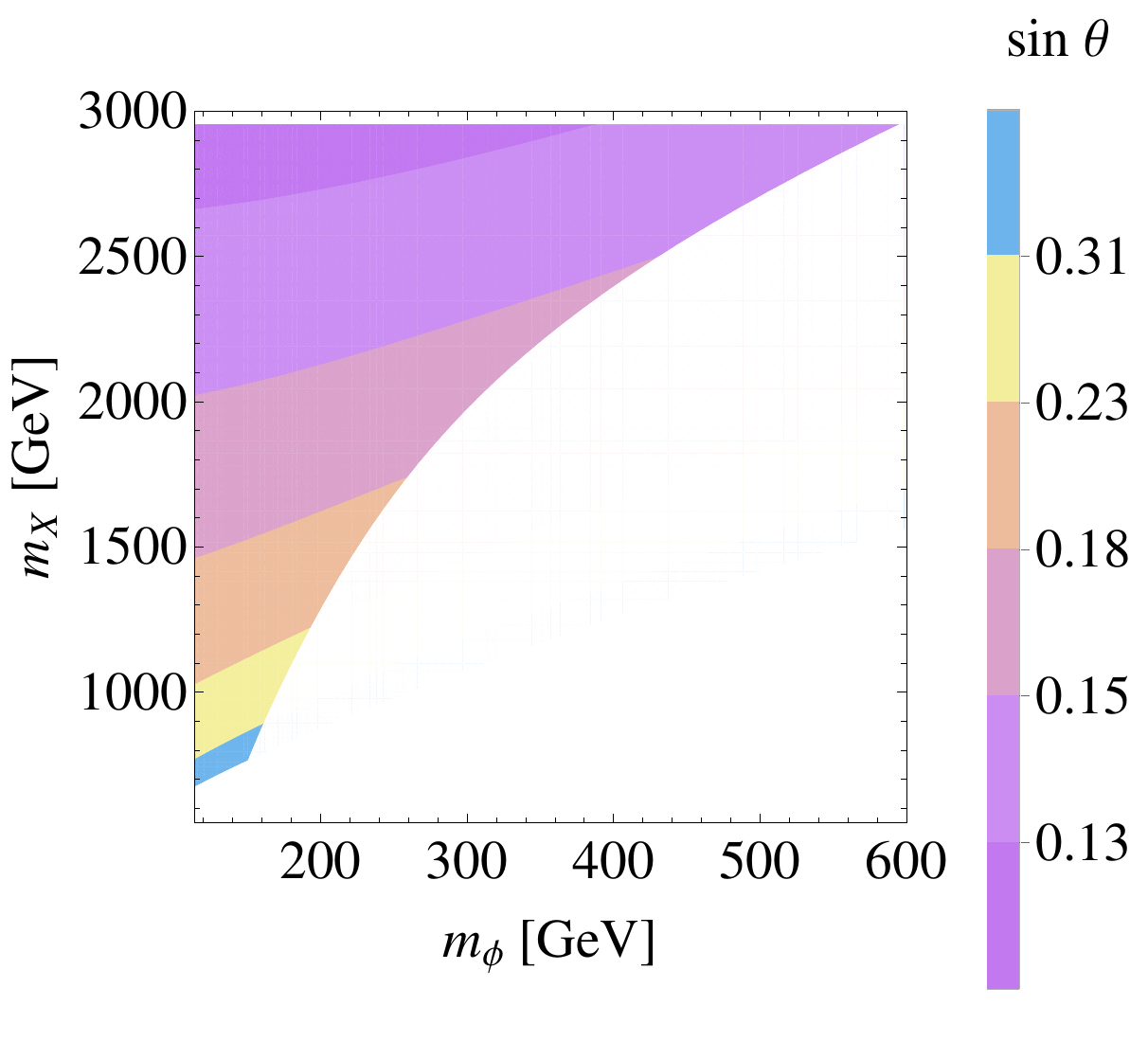}}
\caption{Left (right) panel: The $U(1)_X$ gauge coupling $g_X$
(scalar mixing angle $\sin \theta$)
in the $m_\phi-m_X$ plane for the degenerate mass case.
We impose the LHC bound on $\sin\theta$ and the LUX experiment constraints.}
\label{fig:sinvp1}
\end{figure}

In general, we find moderate values of the dark gauge and the Yukawa couplings.
The Yukawa coupling $y_1$ can be deduced by combining the results
from Fig.~\ref{fig:mngx1} with (\ref{eq:dmmass}) to lie
in the region $1.0 \lesssim y_1\lesssim 2.0$ covered by the corresponding $m_\phi$ and $m_X$ values.
The gauge coupling is shown on the left
panel of Fig.~\ref{fig:sinvp1}.
The behavior of $g_X$ can be understood as follows: a small scalar mixing angle leads to a small portal coupling $\lambda_P$. In order to keep the Higgs mass fixed, this leads to a large \emph{vev} $v_\Phi$ of the scalon in (\ref{eq:vevs}), typically by an order of magnitude larger than the EW scale.
This \emph{vev}, shown on right panel of Fig.~\ref{fig:mngx1}, sets the scale of the DM annihilation processes to the hidden sector.
Therefore, since we know that the DM cross section should be of the order of the EW scale, we need appreciable dark gauge $g_X$ and Yukawa $y_a$ couplings to compensate the large \emph{vev}.
In addition, the CW mechanism favors large gauge couplings, but, at the same time, disfavors large Yukawa couplings, creating some tension between the two.
This is the reason behind the non-monotonous behavior of $g_X$ on the left panel of Fig.~\ref{fig:sinvp1}.

\begin{figure}[h]
\centering
\centerline{\includegraphics[scale=0.55]{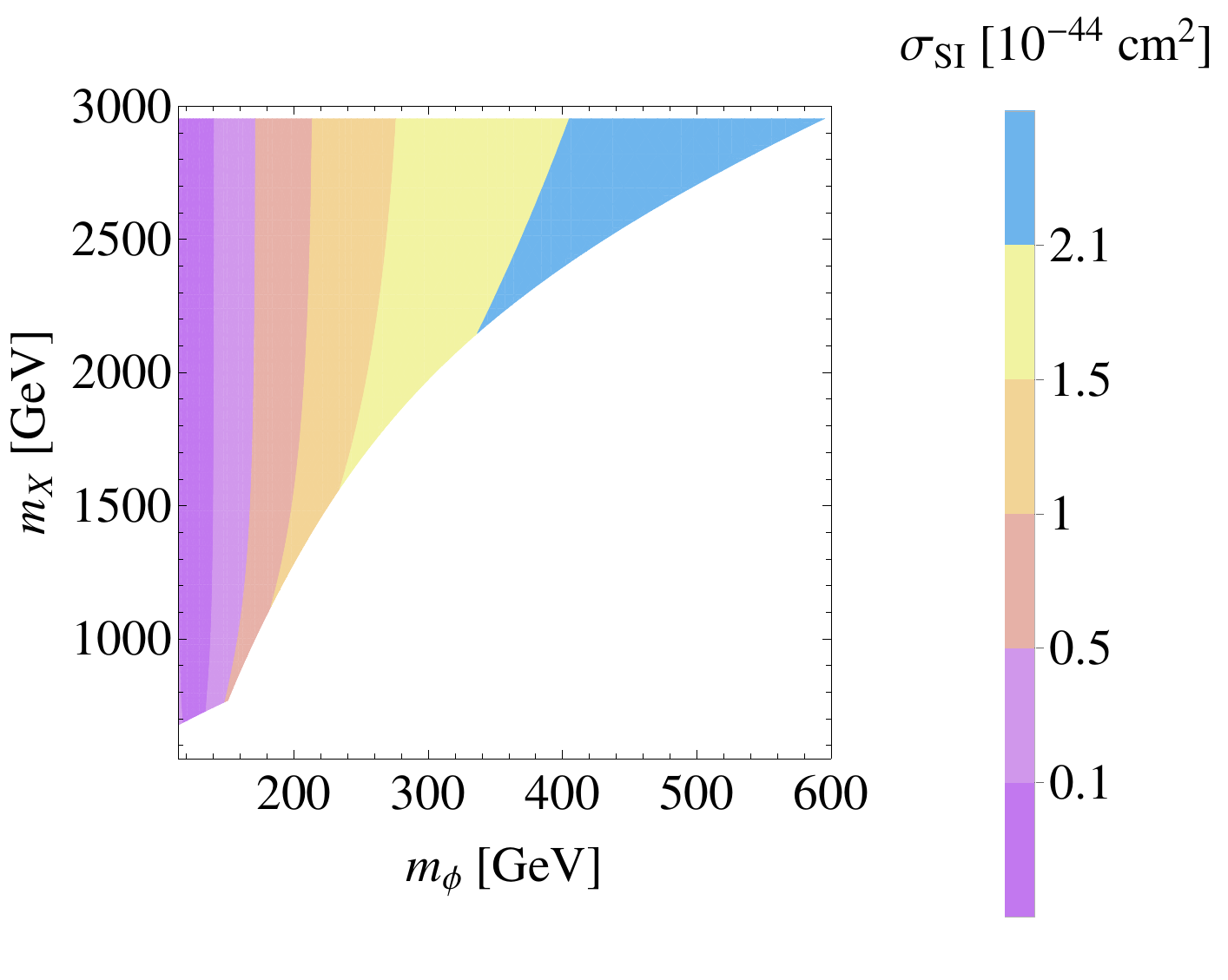}}
\caption{The direct detection cross section $\sigma_{SI}$
in the $m_\phi-m_X$ plane for the degenerate mass case.
We impose the LHC bound on $\sin\theta$ and the LUX experiment constraints.}
\label{fig:dddeg}
\end{figure}

On Fig.~\ref{fig:dddeg} we show the direct detection cross section
$\sigma_{SI}$ as deduced from (\ref{eq:ddmaj}).
The direct detection proceeds through the Higgs portal and is therefore proportional to the mixing angle $\sin^2\theta$.
However, the moderate values of the Yukawa coupling yield
$\sigma_{SI}$ already within the range of the LUX experiment \cite{Akerib:2013tjd}.
The obtained result is largely independent of the $X$-boson mass.
With the future XENON1T and LZ experiments
there is an excellent opportunity to sweep the entire
parameter space of the model except around $m_\phi = m_h$ where $\sigma_{SI}$ drops to zero.

\subsection{The case of non-degenerate masses}

\begin{figure}[h]
\centering
\centerline{\includegraphics[scale=0.53]{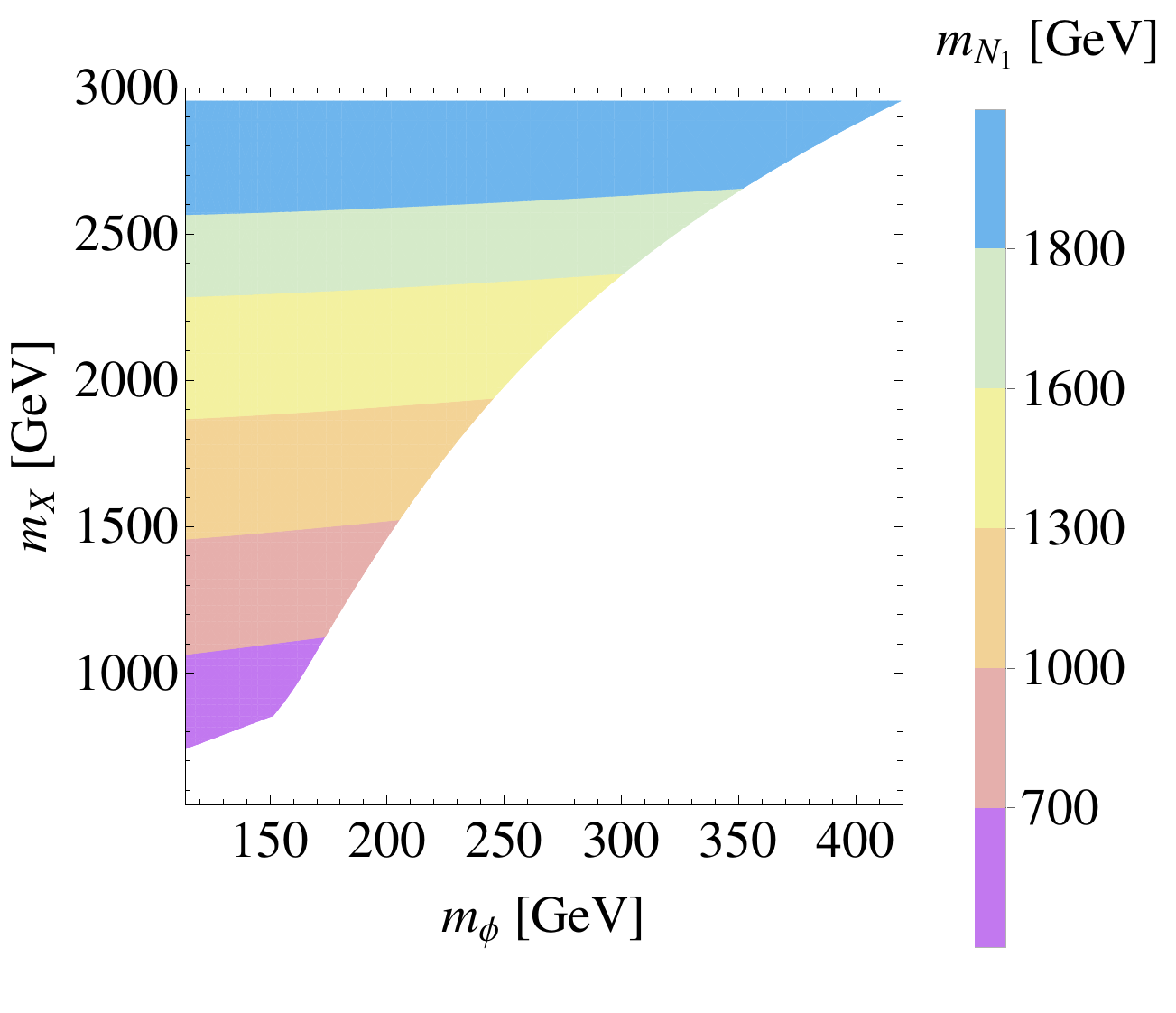}
\hspace{0.25cm}
\includegraphics[scale=0.53]{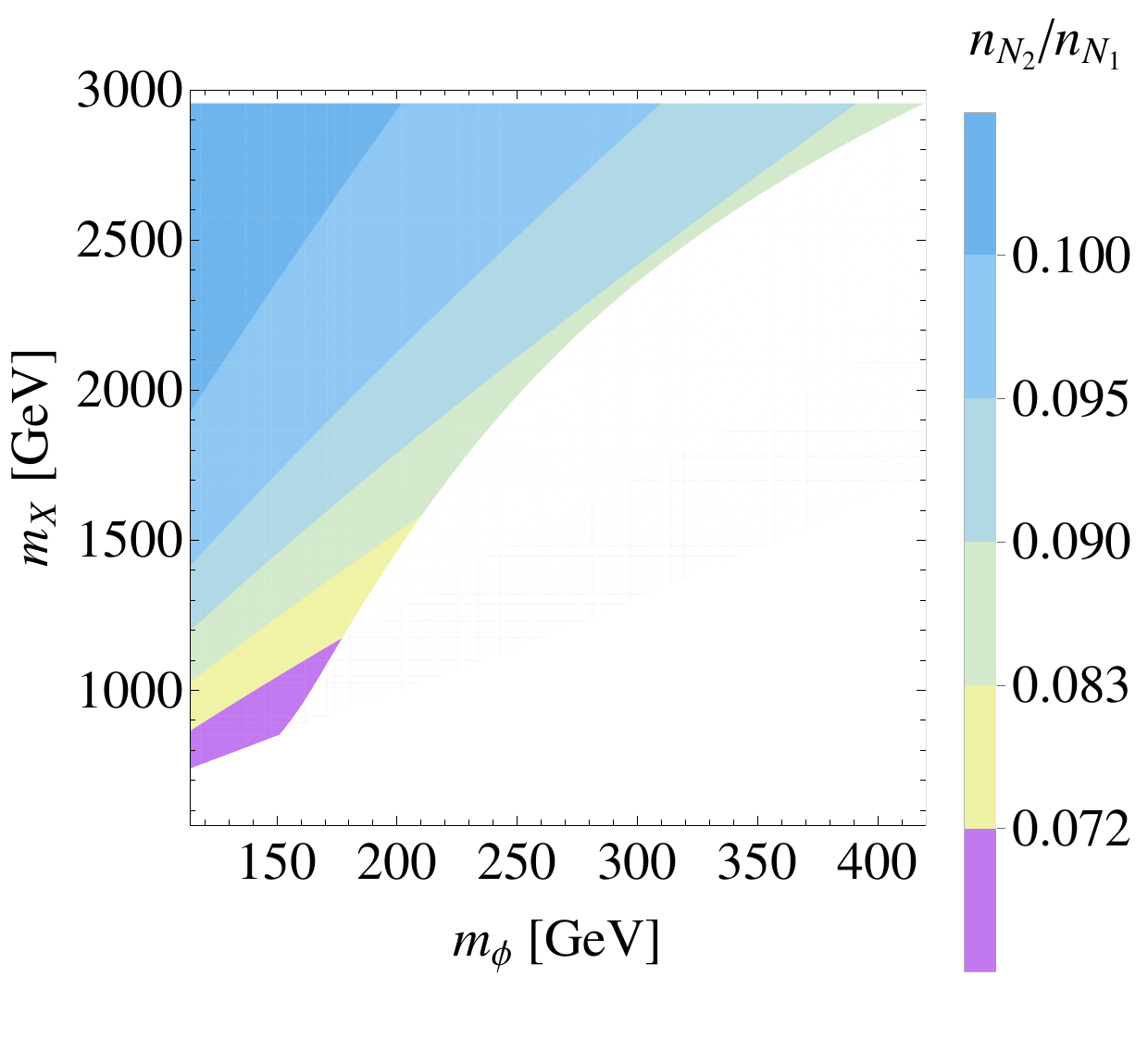}}
\caption{Left (right )panel: Mass of the DM Majorana fermion $m_{N_1}$
(number density ratio $n_{N_2}/n_{N_1}$)
in the $m_\phi-m_X$ plane for $m_{N_2} = 1.5 m_{N_1}$.
We impose the LHC bound on $\sin\theta$ and the LUX experiment constraints.}
\label{fig:mngx2}
\end{figure}

\begin{figure}[h]
\centering
\centerline{\includegraphics[scale=0.53]{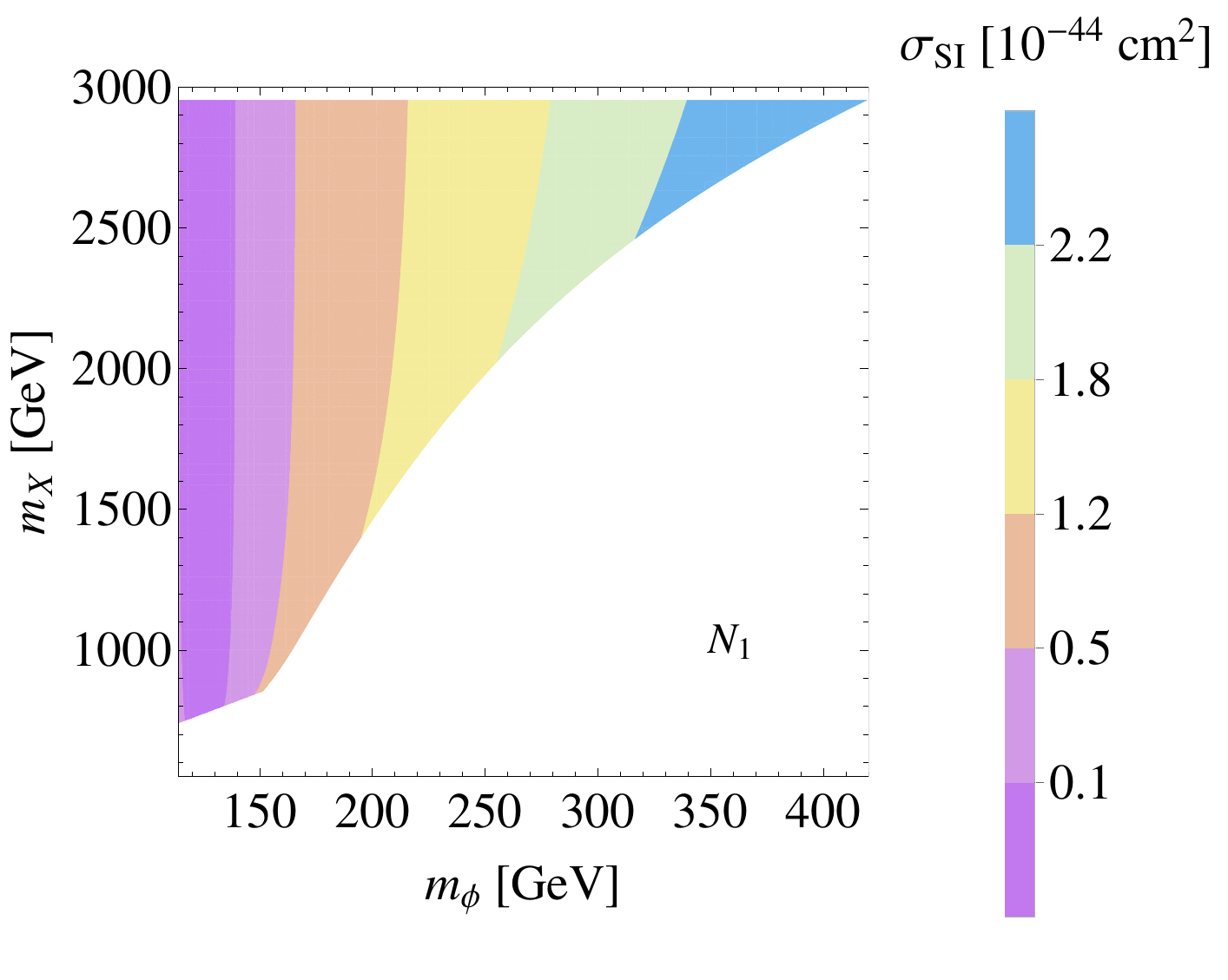}
\hspace{0.25cm}
\includegraphics[scale=0.53]{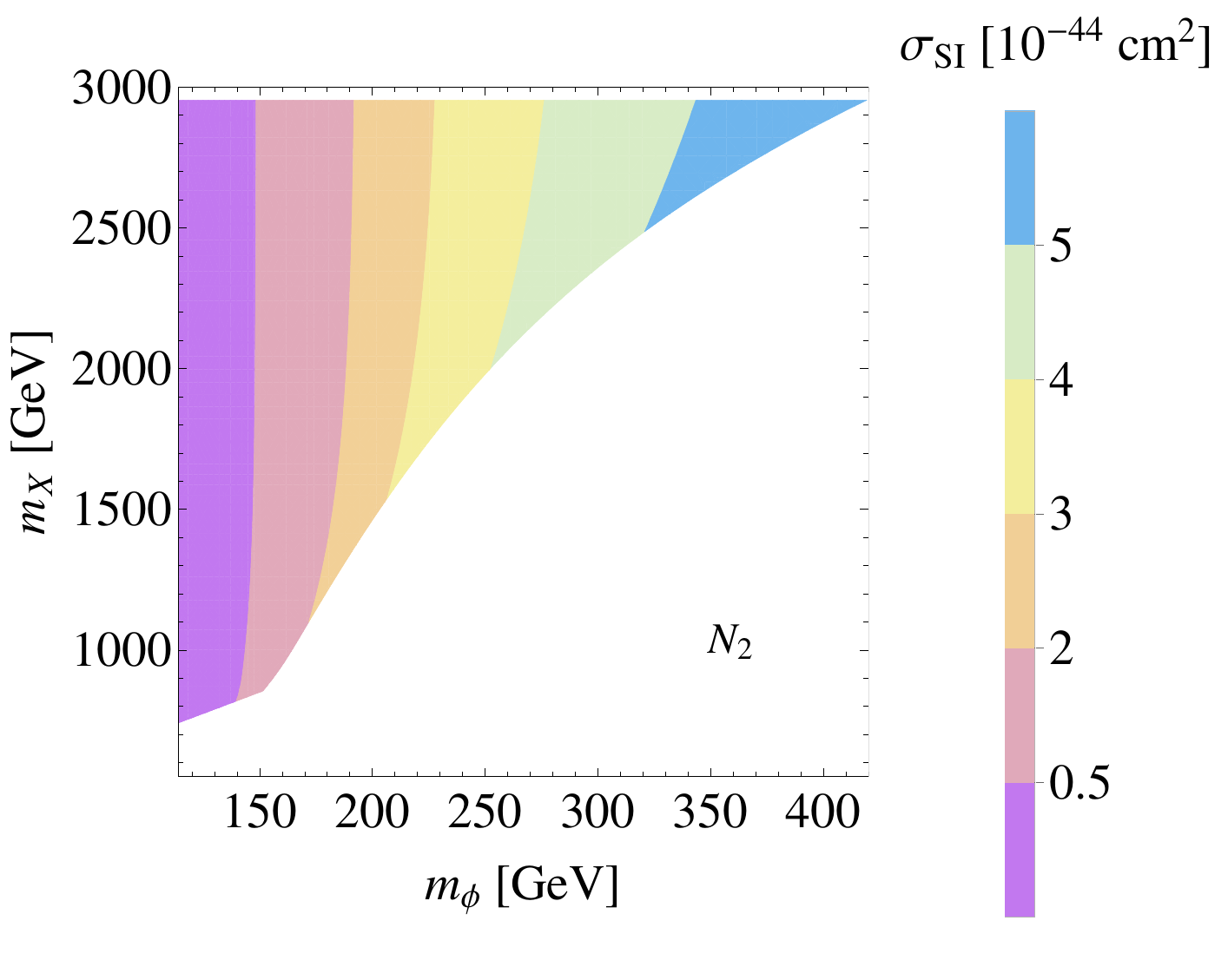}}
\caption{Left (right) panel: Majorana dark
matter $N_1$ ($N_2$) direct detection cross section
in the $m_\phi-m_X$ plane for $m_{N_2} = 1.5 m_{N_1}$.
We impose the LHC bound on $\sin\theta$ and the LUX experiment constraints.}
\label{fig:cross2}
\end{figure}

Here we consider the case where Majorana fermion masses differ,
and for definiteness take
$$m_{N_2} = 1.5 m_{N_1}~,$$
so that the DM Majorana fermions decouple sequentially from
the heat bath.
In this calculation the appearance of the $N_2 N_2 \to N_1 N_1$
annihilation channel has been taken into account.
Our main results are summarized in Figs.~\ref{fig:mngx2} and \ref{fig:cross2}.
On the left panel of Fig.~\ref{fig:mngx2} we see that the lighter DM Majorana fermion mass $m_{N_1}$ spans a similar region as in the case of the degenerate masses.
In fact, the minimum mass of the lighter $N_1$ is roughly the same as in the degenerate case, so that a fixed scalon mass requires a larger dark $X$-boson mass.

The main difference with respect to the degenerate mass scenario is that the DM relic abundance is now mainly saturated by $N_1$.
The $N_2$'s have an extra $N_2 N_2 \to N_1 N_1$
annihilation channel as well as a stronger Yukawa coupling both of which act to decrease the freeze-out abundance
$Y_\infty^2$ according to (\ref{eq:freeze}).
On the right panel of Fig.~\ref{fig:mngx2} we see that the ratio
of the number densities of the heavier $N_2$
with respect to the lighter $N_1$ Majorana
fermion is roughly $10\%$.

On Fig.~\ref{fig:cross2} we plot the direct detection cross section of $N_1$ (left panel) and $N_2$ (right panel).
We have imposed the LUX constraint on cross sections for $N_1$ and $N_2$, appropriately modified to take into account the different number densities. As a direct consequence of the number density dominance of the lighter Majorana fermions $N_1$ the LUX constraints on $N_1$ are more severe in the considered $m_\phi-m_X$ space. Therefore, this is the constraint imposed on both plots of Fig.~\ref{fig:cross2}.
The attractive feature of this scenario is the two distinct signals
in the direct detection experiments.
The numerical values on Fig.~\ref{fig:cross2} indicate that a large portion of the parameter space is testable at future DM searches, such as XENON1T and LZ.

\section{Conclusions}

We have constructed a classically scale invariant model with a dark gauged $U(1)_X$.
The CW mechanism is realized in the dark sector and the scale is transmitted to the
SM through the Higgs portal.
We have introduced a pair of $U(1)_X$-charged chiral fermions.
Classical scale invariance and gauge invariance leave a separate
remnant $Z_2$ symmetry for both Majorana fermions after spontaneous symmetry breaking
making them the DM candidates in our model.
All the masses in the dark sector and the SM sector come from a scale generated dynamically by the CW mechanism.
This makes a connection between the DM mass and the EW scale, accounting for the WIMP miracle.

The model allows for six free parameters, the dark gauge coupling, three couplings in the scalar potential, and two Yukawa couplings in the dark sector.
We have constrained the model by the Higgs mass and \emph{vev} and the observed DM relic abundance.
We have also used LHC constraints on the scalar mixing angle and the LUX experiment results on the direct detection of DM.
The three undetermined parameters were chosen to be the dark scalar and the dark gauge boson masses, and the ratio of Majorana fermion masses.

We have analyzed two possible cases for fixed ratio of the Majorana fermions masses, the case of equal masses $m_{N_2}=m_{N_1}$ and the case with the ratio $m_{N_2}=1.5m_{N_1}$.
In both cases the mass of the Majorana DM can
be from $470$ GeV to a few TeV.
The lower limit is set by the LHC constraints on the mixing angle between the SM and the dark sector scalars, while the upper limit is an estimate from applicability of perturbation theory.
The constraints on the model result in moderate values of the dark gauge and the dark Yukawa couplings.
In the degenerate mass case both Majorana fermions contribute equally to the DM relic abundance, while in the second case the heavier Majorana fermion $N_2$ accounts only for $\sim10\%$ of the number density of $N_1$.

At the LHC the dark sector can be reached through the Higgs portal.
However, due to the large masses in the dark sector the model does not
allow for hidden decay channel of the SM Higgs particle.
The scalon $\phi$ couples to the SM particles through the scalar mixing with a suppression factor given by the mixing angle $\sin\theta$.
LHC data excludes a region of $\sin\theta>0.37$ ruling out the dark $X$-boson masses lower then
$680$ GeV and Majorana DM masses lower then $470$ GeV.
The next LHC run will further test a part of the parameter space of the model.

The direct detection experiments offer the best prospects to test this model.
The interaction with the nucleus proceeds through the Higgs portal, and the key role is played by the dark scalar mass and the Yukawa couplings in the dark sector.
Due to the considerable values of the Yukawa couplings, the spin independent cross section with the nucleus is of the order of $10^{-44}$ cm$^2$ which has been reached in the LUX experiment for the DM mass range obtained in this model.
While the latest LUX constraints already exclude a substantial region of the parameter space, the planned XENON1T and LZ experiments will be able to sweep a majority of the parameter space.
A unique signature for this model would be
two distinctive signals in the direct detection experiments due to the presence of two non-degenerate Majorana DM candidates.

\subsection*{Acknowledgments}
BR would like to thank Hiren Patel, Pavel Fileviez Perez and Sebastian Ohmer for helpful discussions.
This work has been supported in part by the Croatian Science Foundation under the project number 8799.
BR acknowledges the support from the Alexander von Humboldt Foundation.

\end{document}